\documentstyle[preprint,eqsecnum,aps]{revtex}

\begin{document}
\draft

\preprint{SLAC-PUB-6483}

\title{
Charge form factors of pseudoscalar mesons
}
\author{Felix Schlumpf}
\address{
Stanford Linear Accelerator Center\\
Stanford University, Stanford, California 94309
}
\date{\today}
\maketitle

\begin{abstract}
The elastic charge form factors of the pion, kaon, D- and B-mesons are
calculated within a relativistic constituent quark model formulated on the
light-cone.  Our parameter free predictions agree well with the available
data.  The results are approximately
independent of the assumed form of the
light-cone wave function, in particular they do not depend on the high
momentum tail of the wave function for energies accessible to
present experiments.
\end{abstract}


\narrowtext

\section{Introduction}\label{sec:1}

There has recently been renewed interest in the charge form factors of
pseudoscalar mesons because CEBAF is planning experiments
to measure independently the pion (E-93-021) and kaon (E-93-018)
charge form factors in the range of $Q^2<3$ GeV$^2$. On the
other hand, theoretical predictions of the kaon form factor given
in Refs.~\cite{ji90,szcz93,buck94} differ significantly. It is
therefore reasonable to analyze the kaon form factor, and in general the
pseudoscalar form factors, in a different model.

Since an {\em ab-initio} QCD calculation for hadronic wave functions is
currently not feasible, it is important to have QCD-inspired models which
realistically describe hadron static properties. Such a model has been
described in Refs.~\cite{tere76,chun88} and many observables have been
calculated within this model~\cite{jaus90,jaus91,jaus92} and found
in good agreement with experiment.

In this paper we extend the analysis of Ref.~\cite{chun88} to
pseudoscalar mesons in which the constituent quarks have
unequal masses. The main uncertainty in this model is due to the
uncertainty in the undetermined momentum wave function. We find
that the charge form factors are approximately insensitive for
a large class of wave functions. This is a generalization of
the finding in Ref.~\cite{fred92} which studies the connection between
the radius and the decay constant of the pion.

In Ref.~\cite{jaus91} the charge form factors for the pion and kaon are
given in the parameterization $F(Q^2)=F(0)/(1-Q^2/\Lambda_1^2
-Q^4/\Lambda_2^4)$. Unfortunately, this specific representation is only
valid for $Q^2<0.25$ GeV$^2$ for the pion and $Q^2<0.75$ GeV$^2$ for the
kaon. This paper presents the full calculation.

In the next section we describe the model and give the formula
for the charge form factor. In Sec.~III the numerical results
are presented and compared with available experimental data
and other theoretical calculations. Finally, we conclude this
paper in Sec.~IV with a summary.

\section{Relativistic constituent quark model}\label{sec:2}

The light-cone constituent quark model given in Ref.~\cite{chun88}
provides a framework for
representing the general structure of the two-quark wave functions for
mesons.  The wave function is constructed as the product of a
momentum wave function, which is spherically symmetric and invariant under
permutations, and a spin-isospin wave function, which is uniquely
determined by symmetry requirements.  A Wigner (Melosh~\cite{melo74})
rotation is applied to the spinors, so that the wave function of the meson
is an eigenfunction of $J^2$ and $J_z$~\cite{coes82}. To represent the range
of uncertainty in the possible form of the momentum wave function we choose
two simple functions of $p^2$; a
gaussian or harmonic oscillator (HO) wave function and a power-law wave
function:
\begin{eqnarray}
\psi_{\rm HO}(p^2)&=&N_{\rm HO}\exp(-p^2/2\beta^2) ,
\label{eq:ho} \\
\psi_{\rm Power}(p^2)&=&N_{\rm Power}(1+p^2/\beta^2)^{-s} ,
\label{eq:power}
\end{eqnarray}
where $\beta$ sets the scale of the nucleon size, and $p$ is the
three-vector on the light-cone defined below. This may seem
quite arbitrary,
but as we will show below, the form factors are essentially
independent of the shape of the wave function for $s \geq 2$.

Since the four-momentum $Q=p'-p$ is spacelike it is always possible to
orient the axes in such a manner that $Q^+=0$. The charge form factor
$F(Q^2)$ of the pseudoscalar meson is then given by the matrix element
($Q^2=-q^2$)
\begin{equation}
F(Q^2)=\langle p+q \vert J^+(0) \vert p \rangle .
\end{equation}
The general formula for the pseudoscalar meson is then
\begin{equation}
F_{\bar q Q}(Q^2)=e_{\bar q} I(m_{\bar q},m_Q,\beta_{\bar qQ},Q^2)+
   e_Q I(m_Q,m_{\bar q},\beta_{\bar qQ},Q^2) ,
	\label{eq:ff}
\end{equation}
with~\cite{lc}
\begin{eqnarray*}
I(m_1,m_2,\beta_{\bar qQ},Q^2) & =  & \int d^3p \left(
\frac{E_1' E_2' M_0}{E_1 E_2 M_0'} \right)^{1/2}
\psi^\dagger(p')\psi(p) \times \\
 &  & \frac{p_\perp \cdot p_\perp' + (\xi m_2+(1-\xi)m_1)^2}
 {\xi(1-\xi)\sqrt{M_0^2-(m_1-m_2)^2}\sqrt{M_0^{'2}-(m_1-m_2)^2}} .
\end{eqnarray*}
 The relative three-momentum $(p_\perp,p_z)$ in this formula is defined as
\begin{equation}
\xi=p_1^+/P^+ , p_\perp=p_{1\perp}-\xi P_\perp ,
p_z=\xi E_2-(1-\xi)E_1 ,
\end{equation}
where $P$ and $p_1$ are the four-momentum of the meson and the first
constituent quark, respectively. The energy $E_i$ is given by
$E_i^2=p_\perp^2+p_z^2+m_i^2$, the invariant mass $M_0$ is given by
$M_0=E_1+E_2$, and $p'_\perp=p_\perp-(1-\xi)q_\perp$.

The wave function is normalized as $\int d^3p |\psi|^2 =1$ so that $F(0)$
gives the charge of the particle, because $I(m_1,m_2,\beta_{\bar
qQ},0)=1$ and $F(0)=e_{\bar q}+e_Q$.
The radius of the particle is defined by $r^2 = -6
dF(Q^2)/dQ^2\vert_{Q^2=0}$. The expression for the decay constant is
given in Eq.~(3.3) of Ref.~\cite{jaus91}.

\section{Numerical results}\label{sec:3}

The parameters of the model have to be determined by comparison with
experimental data.  They are the constituent masses $m_q$ of the quarks and
the confinement scale $1/\beta$ of the bound meson. The parameters used in
this paper are given in Table~\ref{tab:para}. For the $u-, d-, s-$quark
sector the parameters from Ref.~\cite{jaus91} are chosen. These four
parameters have been obtained by fitting the decay constants $f_\pi,
f_\rho, f_K$ and the decay rate for $K^{*+}\to K^+ \gamma$.
With these parameters all the processes described in Ref.~\cite{jaus91}
can be described in addition to the charge form factors discussed
in this paper. The $c-$quark
mass is taken from Ref.~\cite{jaus92} where the decays $D \to Ke\nu$ and
$D \to K^* e\nu$ are analyzed. There is an uncertainty in $\beta$,
which we determined by fitting the decay constant $f_D$ as given by
recent lattice calculations~\cite{bern93}. The $b-$quark mass is taken from
recent nonrelativistic lattice calculations~\cite{lepa93}.
Again we fit $\beta$
to obtain the decay constant $f_B$ obtained in Ref.~\cite{bern93}.
Table~\ref{tab:para} also shows the $\beta$s for the power-law wave
function in Eq.~\ref{eq:power} with $s=2$, which are determined to yield
the same decay constants as for the  harmonic oscillator wave function.
The weak decay constants for the $D$ and $B$ mesons for different values
of the parameter $\beta$ are given in Table~\ref{tab:wdc}. There is a
strong dependence on $\beta$ so that our model does not restrict the
decay constants of the $D$ and $B$ mesons very well.

The weak decay constants and the charge radii are given in
Table~\ref{tab:static} and compared with data.
All the radii agree with experiment were available.
We also achieve the result $f_B\approx f_D$ given in Ref.~\cite{nari94}.
This differs from heavy quark effective theory which
predicts $f_D/f_B = \sqrt{M_B/M_D}=1.68$. It has already been noted in
Ref.~\cite{clos94} that the Wigner rotation of the spin lowers
this ratio.

Figures \ref{fig:1} -- \ref{fig:6} show the charge form factors of the
pion, kaon, D- and B-mesons, respectively, together with experimental
data  where available. The results for the pion are compared in
Fig.~\ref{fig:1} with data for low $Q^2$~\cite{amen84} and in
Fig.~\ref{fig:2} with data for high $Q^2$~\cite{bebe78}. The solid line
represents the HO wave function in Eq.~\ref{eq:ho} which fits the data
best for both low and high momentum transfer. The other lines give the
result for the power-law wave function in Eq.~\ref{eq:power} with $s=1$
(dotted line), $s=2$ (dashed line), and $s=3$ (dot-dashed line). The
power-law wave function for $s \geq 2$ can describe the data as well. The
result is approximately
independent of the wave function for the experimental
accessible $Q^2$-region~\cite{high}.
This wave function independence is even better
fulfilled for the heavy quark sector as seen in Fig.~\ref{fig:5} and
\ref{fig:6}. This is astonishing since for baryons it was  found in
Ref.~\cite{schl93} that the high momentum tail of the
wave function is important for $Q^2>2$
GeV$^2$. On the other hand, Ref.~\cite{schl94} describes that static
properties are exactly independent of the wave function even for the
light quark sector. The result of Fig.~\ref{fig:1} is expected from
Ref.~\cite{fred92} which confirms the approximate
relation $r_\pi = \text{const}/f_\pi$
for different wave functions.

Figure~\ref{fig:2} shows that the high momentum tail of the wave function
does not matter for energies accessible to present experiments. Important
is the parameter $\beta$ which determines the overall shape of the wave
function. This has independently been found by Jean~\cite{jean}.
Indeed all the different curves in Fig.~4 of
Ref.~\cite{card94} can be obtained with a HO wave function with
fixed quark mass and different $\beta$s.

The results for the kaon are compared in Fig.~\ref{fig:3} with data for
low $Q^2$~\cite{amen86} and shown in Fig.~\ref{fig:4} for high $Q^2$.
Again, the solid and the dashed lines give the result for the HO and
power-law wave function for $s=2$, respectively. This should represent
the range of uncertainty in the possible form of the momentum wave
function. Our prediction is compared with the calculation in
Ref.~\cite{buck94} (dotted line) and in Ref.~\cite{szcz93} (dot-dashed
line). While the former curve nearly coincides with our power-law wave
function result, the latter calculation gives a much smaller charge form
factor. The CEBAF experiment (E-93-018) may be able to distinguish
between these different predictions.

Our  formalism is suited for both light and heavy quarks, so that we also
present the charge form factors for the D- and B-mesons in
Fig.~\ref{fig:5} and \ref{fig:6}. Note the different scale in the
momentum transfer. The wave function dependence for the D- and B-mesons is
very small even for very high momentum transfer.

A main emphasis of this paper is to choose the same parameters as
Ref.~\cite{jaus91}, so that all the precise results obtained in that
paper are preserved. For instance, the explicit calculation of the weak
$K\to\pi$ transition form factor gives $F(0)=0.965$ and $r^2=0.318$
fm$^2$ for the HO wave function~\cite{jaus91}. For $K_{e3}$ decay, one may
neglect the electron mass in the calculation of the rate, which then
depends only on the form factor $F(Q^2)$.
The corresponding value $V_{us}$ of the
KM quark-mixing matrix is $V_{us}=0.2199\pm 0.0017$~\cite{jaus91}.
The error for $V_{us}$ originates in the uncertainty of the
strange quark mass in the constituent quark model.

\section{Conclusion}\label{sec:4}

We  extend the simple relativistic quark  model described in
Ref.~\cite{chun88,jaus90,jaus91} to investigate the charge form factors
of the pseudoscalar mesons. Most of the
parameters are fixed by these previous
studies of the model. We notice an approximate
independence of the momentum
wave function for a large $Q^2$-range. The charge form factors do not
depend on the high momentum tail of the wave function, but rather on the
confinement scale $1/\beta$.

The validity of this model has also been tested in other systems such as
mesons having nonzero angular momentum~\cite{jaus91} and
baryons~\cite{schl93}. In conclusion, this model provides a remarkably
good description of the electroweak properties of hadrons, although it
is both conceptually and computationally simple.

\acknowledgments

It is a pleasure to thank S.~Brodsky, F.~Coester, H.-C.~Jean, and
G.~Salm\`e for helpful discussions.
This work was supported in part by the Schweizerischer Nationalfonds and
in part by the Department of Energy, contract DE-AC03-76SF00515.

\begin{table}
\caption{Table of the parameters of the relativistic constituent
quark model.
\label{tab:para}}
\begin{tabular}{cddd}
$q$ & $m_q$ (GeV) & $\beta_{u \bar q}^{\rm HO}$ (GeV) &
$\beta_{u \bar q}^{\rm Power}$ (GeV) \\
\tableline
$u,d$ & 0.25 & 0.3194 & 0.335 \\
$s$   & 0.37 & 0.395 & 0.41 \\
$c$   & 1.85 & 0.49 & 0.50 \\
$b$   & 4.7  & 0.55 & 0.54 \\
\end{tabular}
\end{table}

\begin{table}
\caption{The weak decay constants of the $D$ and $B$ mesons for
different values of the parameter $\beta$ for the HO wave
function.
\label{tab:wdc}}
\begin{tabular}{crrrr}
$\beta$ (GeV) & 0.4 & 0.5 & 0.6 & 0.7 \\
\tableline
$f_D$ (MeV) & 118 & 149 & 178 & 205 \\
$f_B$ (MeV) & 89 & 117 & 147 & 177 \\
\end{tabular}
\end{table}

\begin{table}
\caption{Charge radii and weak decay constants for the
pseudoscalar mesons. The parameters have been chosen  in such a way that
both HO and power-law wave function give the same weak decay constant. The
charge radii are given for the HO wave function.
\label{tab:static}}
\begin{tabular}{ccc}
Observable & Expt. & Calculation \\
\tableline
$f_\pi$ (MeV) & $92.4 \pm 0.2$ \protect\cite{hols90} & 92.4 \\
$f_K$ (MeV) & $113.4 \pm 1.1$ \protect\cite{hols90} & 113.4 \\
$f_D$ (MeV) & $<219$ \protect\cite{pdg} & 146 \\
$f_B$ (MeV) & -- & 132 \\
\tableline
$r_\pi^2$ (fm$^2$) & $0.432 \pm 0.016$ \protect\cite{amen84}& 0.449 \\
 & $0.44 \pm 0.03$ \protect\cite{dall82} & \\
$r_{K^+}^2$ (fm$^2$) & $0.34 \pm 0.05$ \protect\cite{amen86} & 0.327 \\
$r_{K^0}^2$ (fm$^2$) & $-0.054 \pm 0.101$ \protect\cite{amen86} & 0.000 \\
$r_D^2$ (fm$^2$) & -- & 0.024 \\
$r_B^2$ (fm$^2$) & -- & 0.005 \\
\end{tabular}
\end{table}

\begin{figure}
\caption{The square of the pion charge form factor for low values of
$Q^2$ compared  with data~\protect\cite{amen84}. The solid line
represents the HO wave function, the other lines give the result for the
power-law wave function with $s=1$ (dotted), $s=2$ (dashed), and $s=3$
(dot-dashed).
\label{fig:1}}
\end{figure}

\begin{figure}
\caption{The charge form factor for the pion compared with data taken
from Ref.~\protect\cite{bebe78}. The same line code as in
Fig.~\protect\ref{fig:1} is used.
\label{fig:2}}
\end{figure}

\begin{figure}
\caption{The square of the kaon charge form factor for low values of
$Q^2$ compared  with data~\protect\cite{amen86}. Solid line, HO wave
function; dashed line, power-law wave function with $s=2$; and
the dot-dashed line is taken from Ref.~\protect\cite{szcz93}.
\label{fig:3}}
\end{figure}

\begin{figure}
\caption{The charge form factor for the $K^+$. The same line code as in
Fig.~\protect\ref{fig:3} is used. In addition, the dotted
line is taken from Ref.~\protect\cite{buck94}.
\label{fig:4}}
\end{figure}

\begin{figure}
\caption{The charge form factor for the $D^\pm, B^\pm$. Solid line,
HO wave function; dashed line, power-law wave function with $s=2$.
\label{fig:5}}
\end{figure}

\begin{figure}
\caption{The charge form factor for the $D^+, B^+$. The same line
code as in Fig.~\protect\ref{fig:5} is used.
\label{fig:6}}
\end{figure}


\begin{thebibliography}{10}
\bibitem{lc}
To write the integral in usual light-cone variables one can replace
$dk_3$ by $d\xi E_1 E_2 / \xi (1-\xi) M_0$.
\bibitem{high}
At much higher momentum transfer the charge form factor has a different
falloff for the different wave functions. We do not attempt to
relate this model to the perturbative limit of
QCD~\cite{lepa80}.
\end{thebibliography}
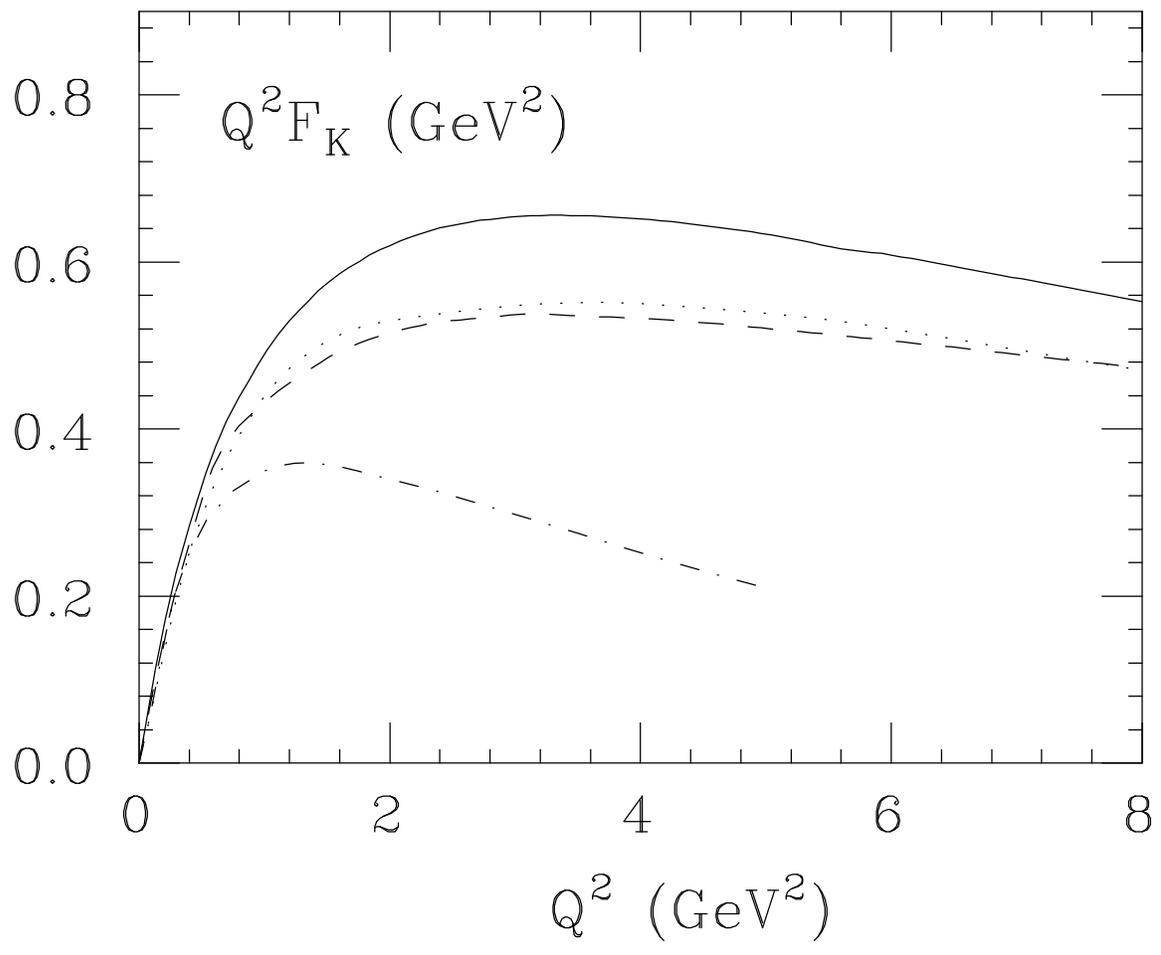
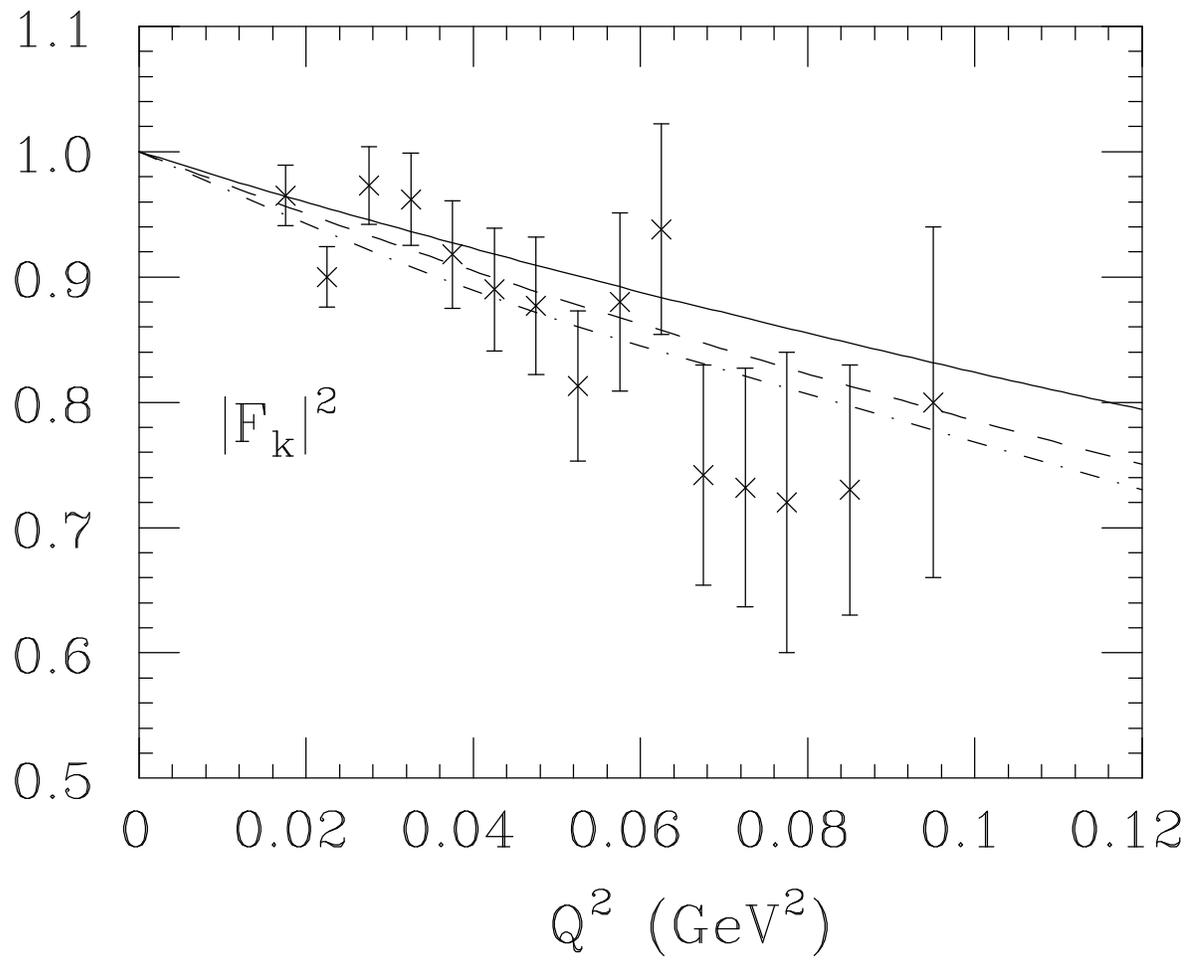
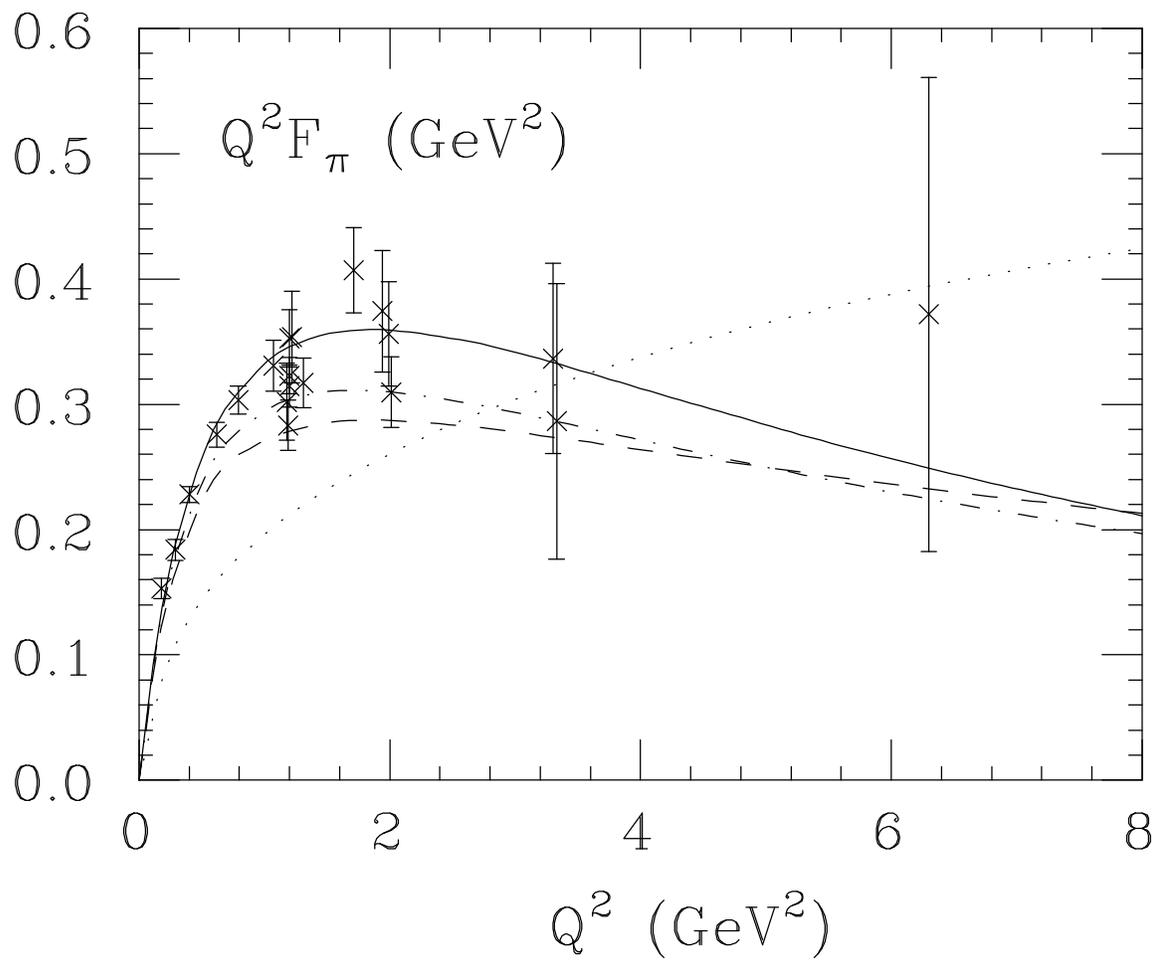
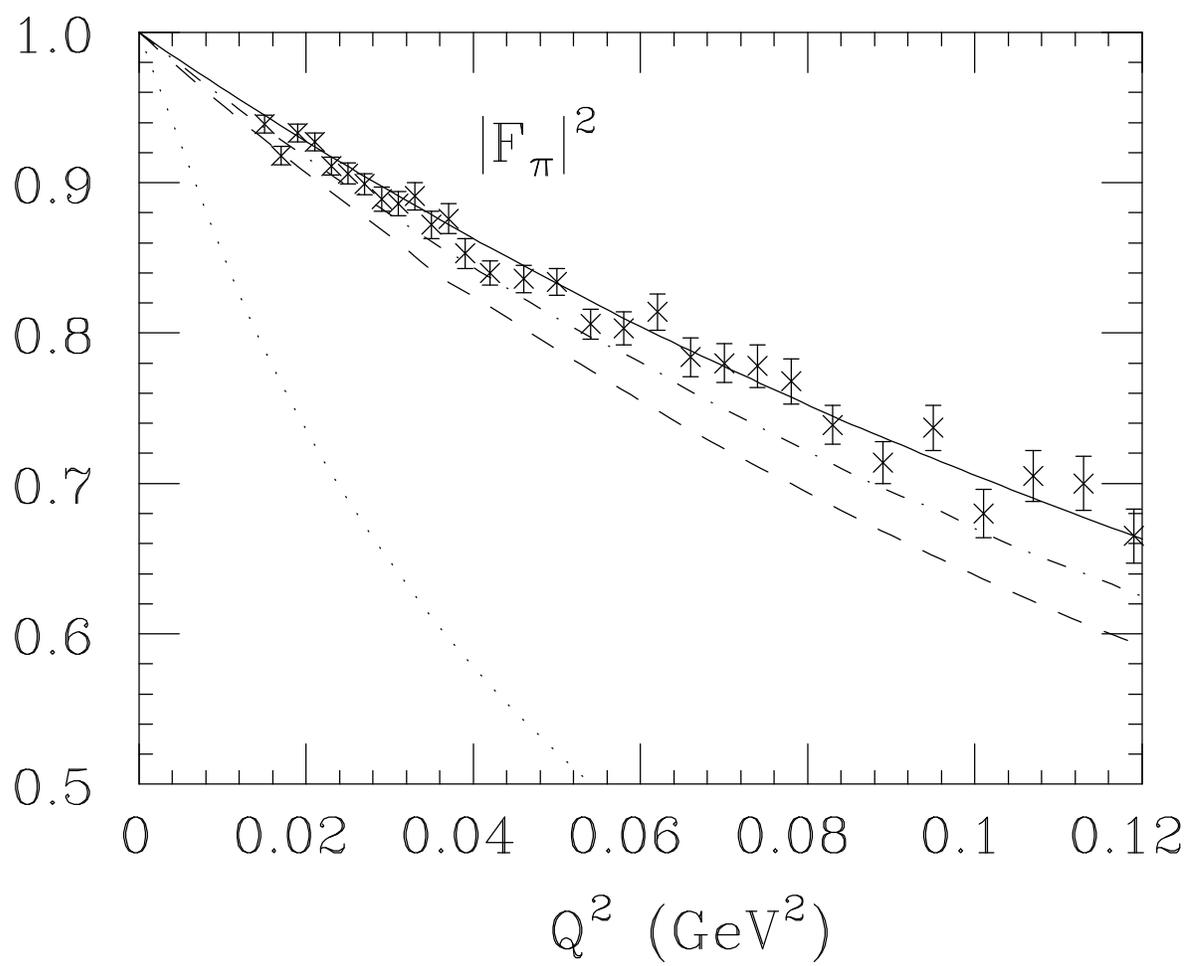
\end{document}